\documentclass[reprint,twocolumn,superscriptaddress,preprintnumbers,amsmath,amssymb,aps,prd,floatfix]{revtex4-2}

\usepackage[utf8]{inputenc}

\usepackage{mathtools}
\usepackage{amsfonts}
\usepackage{mathrsfs}
\usepackage{bm}
\usepackage[normalem]{ulem}
\usepackage{bbold}
\usepackage{braket}
\usepackage{slashed}
\usepackage{dsfont}
\usepackage{float}
\usepackage{dcolumn}% Align table columns on decimal point
\usepackage{amssymb,amsmath}
\usepackage{titlesec}

\usepackage{graphicx}
\usepackage{color}
\usepackage{array}

\usepackage{placeins}
\usepackage{booktabs}
\usepackage{xspace}
\usepackage{hyperref}
\usepackage[nameinlink]{cleveref}
\usepackage{bookmark}
\usepackage{units}
\usepackage{ulem}

\setkeys{Gin}{width=0.48\textwidth}

\usepackage{booktabs}
\usepackage{multirow}

\newcolumntype{C}{>{$}c<{$}}
\AtBeginDocument{
	\heavyrulewidth=.08em
	\lightrulewidth=.05em
	\cmidrulewidth=.03em
	\belowrulesep=.65ex
	\belowbottomsep=0pt
	\aboverulesep=.4ex
	\abovetopsep=0pt
	\cmidrulesep=\doublerulesep
	\cmidrulekern=.5em
	\defaultaddspace=.5em
}

\graphicspath{{./figures/}}
\usepackage{xifthen}
\usepackage{xcolor}

\newcommand{\gettitle}{Classical elasticity meets quantum complexity: \\ A connection from the holographic lens}

\hypersetup{
	colorlinks,
	linkcolor={blue!75!black},
	citecolor={blue!75!black},
	urlcolor={blue!75!black}, 
	pdftitle={\gettitle},
	pdfauthor={},
	pdfkeywords={}
	{} 
	bookmarksopen=true,
	bookmarksopenlevel=2,
	bookmarksnumbered=true
}
\begin{document}

\title{\gettitle}

\author{Yuanceng Xu}
\affiliation{Institute of Theoretical Physics, School of Physics, Dalian University of Technology,
Dalian 116024, China}
\author{Wei-Jia Li}
\thanks{weijiali@dlut.edu.cn (corresponding author)}
\affiliation{Institute of Theoretical Physics, School of Physics, Dalian University of Technology,
Dalian 116024, China}

\date{\today}

\begin{abstract}
In this work, we explore the effects of shear deformations in a wide class of holographic amorphous solids. It is found that both the shear stress and the complexity of formation grow with the increase of the shear strain. Notably, in the regime of very large shear, they exhibit coordinated behavior and adhere to a universal scaling relation, uncovering a surprising connection between two seemingly unrelated aspects of amorphous systems. Furthermore, our findings also provide a counterexample to the previous understanding that the complexity scales linearly with the Bekenstein-Hawking entropy for large static black holes.
\end{abstract}
\maketitle

%######################################################################%
\section{Introduction}
Elasticity, which describes the response of solid-like materials to mechanical deformations, has long been studied within the framework of classical phenomenologies. Conventionally, the behavior of solids under infinitesimal deformations can be well described by the theory of linear elasticity \cite{landau7}. While the elastic behavior of crystalline solids arises from the long-range translational order, the emergence of rigidity and the associated elastic response in amorphous systems have a more complicated and still not fully understood origin. On the other hand, understanding the mechanical properties of amorphous solids, especially under finite deformations, is crucial from the perspective of materials science and engineering. This also encompasses a variety of active topics in soft matter physics, such as strain softening and hardening, plasticity, yielding, and jamming transitions, etc. \cite{Lyulin2005StrainSA,PhysRevX.6.011005,RevModPhys.89.035005,RevModPhys.82.789}. However, aside from the elastoplastic models, a unified description of the non-linear behavior of complex materials remains elusive \cite{ZAMM:ZAMM19850650903,RevModPhys.90.045006}. Recent studies of amorphous systems have shown that quantum theories and related computational methods can provide important insights into a deeper understanding of microscopic mechanisms and physical properties in certain situations \cite{Castelnovo:2010,10.1007/BFb0104822,Novikov:2013}. This also raises the need for new theoretical frameworks that connect the physical quantities traditionally described by classical phenomenologies to those in quantum mechanics.

Over the past few decades, the string-inspired AdS/CFT correspondence (also known as holographic duality) has been applied to gain insights into various quantum many-body systems with strong correlations in nature. These systems range from QCD matter in high-energy physics to exotic phases in condensed matter physics. The holographic duality enables us to tackle challenging problems in  quantum many-body systems by using classical theories of gravity in one higher dimension. Nowadays, it has become an important tool for exploring the physics of strongly correlated regimes where conventional perturbative methods break down. For reviews of developments in these fields, see \cite{Casalderrey-Solana:2011dxg,Ammon_Erdmenger_2015,Zaanen_Liu_Sun_Schalm_2015,Hartnoll:2016apf,Baggioli:2019rr}. 

In this work, we explore the intriguing intersection of non-linear elasticity and quantum complexity in amorphous solids through the lens of holographic duality, with a particular focus on how these two quantities are affected by finite shear deformations. The computations are based on a class of holographic axion models that have been widely applied to modeling viscoelastic materials in previous studies (see \cite{Baggioli:2021xuv} for a detailed review.). With more in-depth research on such models in recent years, it has been found that they share key features with amorphous solids rather than crystalline systems: 
1. Unusual thermodynamic properties, including non-vanishing zero-temperature entropy and non-minimal free energy, indicating that the systems are in multiple metastable states and have the glassy nature \cite{Sachdev:2015efa,Alberte:2017oqx,DeGiuli:2020zew,Facoetti:2019rab}. 
2. Lack of long-range translational order and commensurability effects in these systems \cite{Andrade:2015iyf}. 
3. Enhanced entropy under shear strains, in stark contrast to the decreased entropy observed in crystals \cite{Jin:2003,Pan:2021cux}. In particular, the evolution of entropy under very large shear exhibits a scaling behavior, in agreement with the results from simulations of granular matter close to the jamming transition \cite{Pan:2021cux}. 

This work further unveils a surprising connection between non-linear elasticity and complexity in amorphous systems which has a very universal form: In the large shear regime, the stress and the complexity of formation satisfy a scaling relation 
\begin{eqnarray}\label{scalingintro}
\Delta\mathcal{C}\sim \sigma^{d/(d+1)}.
\end{eqnarray}
It is worth noting that, unlike the scaling behavior of entropy found in \cite{Pan:2021cux}, the power in \eqref{scalingintro} only depends on the spatial dimension $d$ and is insensitive to the details of the system. This finding not only enriches our understanding of the interplay between the mechanical properties of amorphous materials and quantum information but, more importantly, also paves a new avenue for the holographic method to bridge gaps between seemingly disparate fields.

%######################################################################%
%_______________________________________
\section{Holographic model}
%_______________________________________
In this section, we briefly overview the setup of the gravitational model and some key features obtained in previous holographic studies. More technical details are provided in the Appendix. Let us consider the following 4-dimensional holographic action:
\begin{eqnarray}\label{action}
    S=\int d^{4}x \sqrt{-g}\left[R-2\Lambda-2m^2V(X,Z)   \right],
\end{eqnarray}
where $R$ is the Ricci scalar, $\Lambda$ is the negative cosmological constant, $m^2$ is an effective coupling with the dimension of mass squared, $V(X,Z)$ is a general function which we will take to be
\begin{eqnarray}
V(X,Z)=X^MZ^N
\end{eqnarray}
with two parameters $M$ and $N$, and we have adopted the convention $16\pi G_N\equiv 1$. In addition, we have defined $\mathcal{I}^{IJ}\equiv\partial_{\mu}\phi^{I}\partial^{\mu}\phi^{J}$ with 
$X\equiv\frac{1}{2}\text{Tr}\left[\mathcal{I}^{IJ}\right]$, $Z\equiv\det[\mathcal{I}^{IJ}]$ and $I={x,y}$. 

The background values of the Stückelberg scalars (which is also called axions in numerous holographic studies) $\phi^I$ are chosen as 
\begin{eqnarray}   
\bar{\phi}^{I}=M^{I}_{i}x^{i},\quad i=x,y, \label{axions}
\end{eqnarray}
where $M^{I}_{i}$ is a 2$\times$2 symmetric matrix independent of the spatial coordinates $x^i$. Such choice of the scalars' profile endows bulk gravitons with masses while retaining the homogeneity of the background geometry in $x^i$-directions. As a result, it realizes the breaking of the global translations of the boundary system in an economic way \cite{Vegh:2013sk}. Generally, one can take the following profile for the bulk axions,
\begin{eqnarray}\label{shearpro}
M^{I}_{i}=\alpha\begin{pmatrix}
    \sqrt{1+\epsilon^2/4} &  \epsilon/2 \\
    \epsilon/2             &   \sqrt{1+\epsilon^2/4}
\end{pmatrix},
\end{eqnarray}
where $\alpha$ and $\epsilon$  are two dimensionless parameters with the latter one further breaking the rotations on the $x-y$ plane.

Furthermore, it was noted in \cite {Alberte:2017oqx} that one should require that $M+2N>5/2$ to ensure the spontaneous breaking of translational symmetry and rotational symmetry in boundary systems. Then, following the perspective of the effective field theory of elasticity, $\alpha\neq 1$ characterizes the variation of volume(or area in two spatial dimensions), while $\epsilon$ corresponds to shear deformation that preserves volume \cite{Alberte:2018doe}. In the rest of this paper, we will only focus on purely sheared systems, which means setting $\alpha = 1$ and allowing $\epsilon$ to take on finite values.
With this setup, the ansatz for the asymptotic AdS background becomes
\begin{align}
    ds^2=\frac{L^2}{u^2}\left[-f(u)e^{-\chi(u)}dt^2+\frac{1}{f(u)}du^2+\gamma_{ij}(u)dx^{i}dx^j\right],  \label{metic}
\end{align}
where $u$ is the radial coordinate, extending from the boundary ($u=0$) to the black hole horizon ($u=u_h$) defined by $f(u_h)=0$, and $\text {det}\left(\gamma_{ij}\right)=1$. The functions $f$, $\chi$ and $\gamma_{ij}$ can be obtained by solving the Einstein equation numerically. In the following, we set $L\equiv 1$ as in most holographic studies for simplicity.

According to the holographic dictionary, the stress tensor on boundary can be extracted from the asymptotic behavior of $\gamma_{ij}$ that is 
\begin{eqnarray}\label{UVexp}
\gamma_{ij}(u)\sim \gamma_{ij}^{(0)}+\dots+\gamma_{ij}^{(3)}u^3+\dots,
\end{eqnarray}
where $\gamma_{ij}^{(0)} $ corresponds to an external source of the stress tensor operator $T_{ij}$ and $\gamma_{ij}^{(3)}$ is interpreted as its expectation value $\langle T_{ij} \rangle$. For pure shear, one can find  numeric solutions where $\gamma_{xy}$ obeys equation \eqref{UVexp} near the boundary with the coefficients $\gamma_{xy}^{(0)}=0$, and $\gamma_{xy}^{(3)}\neq 0$ dual to a non-zero shear stress denoted by $\sigma\equiv \langle T_{xy}\rangle$. In this case, the shear strain is not supported by the gravity sector but is solely contributed by the profile of the axions, i.e., equation \eqref{shearpro} with a finite $\epsilon$. 

The non-linear stress-strain curve can then be extracted from the dependence of $\gamma_{ij}^{(3)}$ on  $\epsilon$. It was found in \cite{Baggioli:2020qdg} that the shear stress displays a scaling behavior
\begin{eqnarray}\label{stressscaling}
\sigma\sim\epsilon^{\nu},\quad\nu=\frac{3M}{M+2N},
\end{eqnarray}
under sufficiently large strain \footnote{For high enough temperature, another scaling appears for intermediate shear which however disappears at low temperatures.}. Apart from this, it was also revealed that the entropy density grows under shear and exhibits a scaling law
\begin{eqnarray}\label{enscaling}
s\sim \sigma^{{\xi}},\quad 
\xi^{-1}=\frac{3}{2}\left(1+\frac{\nu^2}{9}\right),
\end{eqnarray}
for large shear stress \cite{Pan:2021cux}. This is in contrast with the behavior of crystalline systems where the thermal entropy eventually decreases to zero under shear \cite{Jin:2003}.

%%%%%%%%%%%%%%%%%%%%  
%_______________________________________
\section{Complexity of formation under shear}

\begin{figure}
    \centering
\includegraphics[width=0.8
    \linewidth]{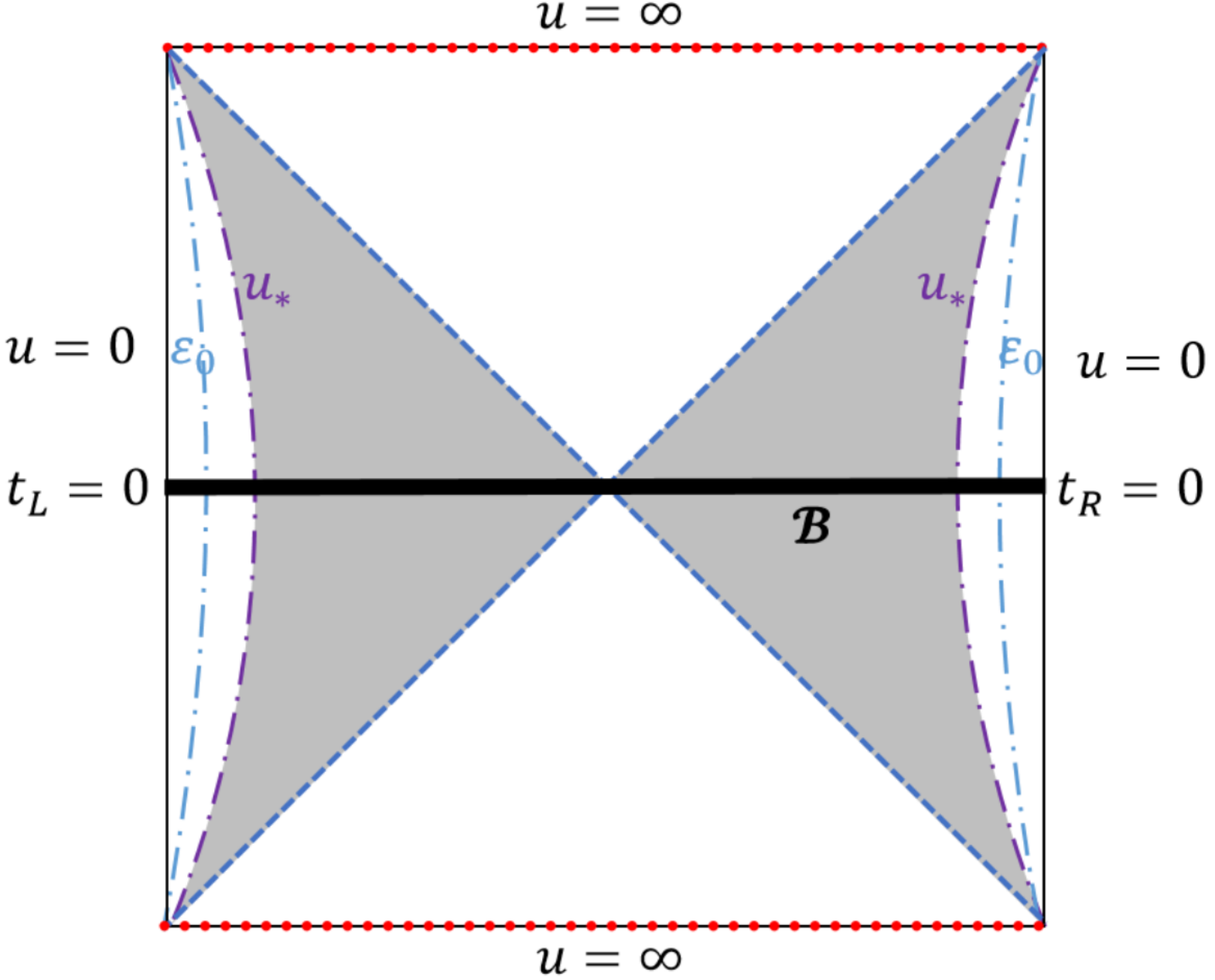}
    \caption{The Penrose diagram of the sheared two-sided black hole: The thick horizontal line denotes the constant time slice $\mathcal{B}$ with maximal volume, connecting the two boundaries at $t_L=t_R=0$ through the Einstein-Rosen bridge. The shaded patches denote the emergent geometry \eqref{scalinggeo} induced by the large shear, which is bounded by the horizon and an emergent scale $u_*$.}\label{ERB}
\end{figure}

Now, we study the effects of shear on the quantum complexity of the system. In holography, the complexity can be computed using the geometry of eternal two-sided black holes, which are dual to a class of entangled states called thermal field double
(TFD) states which can be expressed as \cite{Maldacena:2001kr}
\begin{eqnarray}\label{TFD}
|\text{TFD}\rangle=\frac{1}{Z}\sum_{n}e^{-\frac{E_n}{2T}}e^{-iE_n(t_\text{L}+t_\text{R})}|n\rangle_\text{L}\otimes|n\rangle_\text{R},
\end{eqnarray}
where $t$ is the time, $T$ is the temperature, $E_n$ and $|n\rangle$ are the eigen energy and the associated eigen states, $Z$ is the partition function, $\text{L}$ and $\text{R}$ label the two decoupled entangled subsystems. The complexity=volume (CV) conjecture equates the complexity to the volume of the extremal/maximal constant time slice anchored at boundary times $t_\text{L}$ and $t_\text{R}$ \cite{Susskind:2014rva,Stanford:2014jda}
\begin{eqnarray}
\mathcal{C}=\text{max}\left[16\pi \mathcal{V}(\mathcal{B}) \right].
\end{eqnarray}
where the prefactor $16\pi$ comes from the convention of $16\pi G_N\equiv 1$ and $L\equiv1$ we used. The holographic complexity can, in general, be time-dependent. The focus of this paper is on systems with fixed boundary times that are symmetric with respect to the left and right boundaries, i.e., $t_\text{L}=t_\text{R}=0$. The maximum volume wormhole, given by the $t=0$ slice in the bulk, is represented in the Penrose diagram as a straight line connecting the two boundaries through the bifurcation surface of the horizons. An illustration has been shown in Fig.\ref{ERB}.

\begin{figure}
    \centering
\includegraphics[width=0.9
    \linewidth]{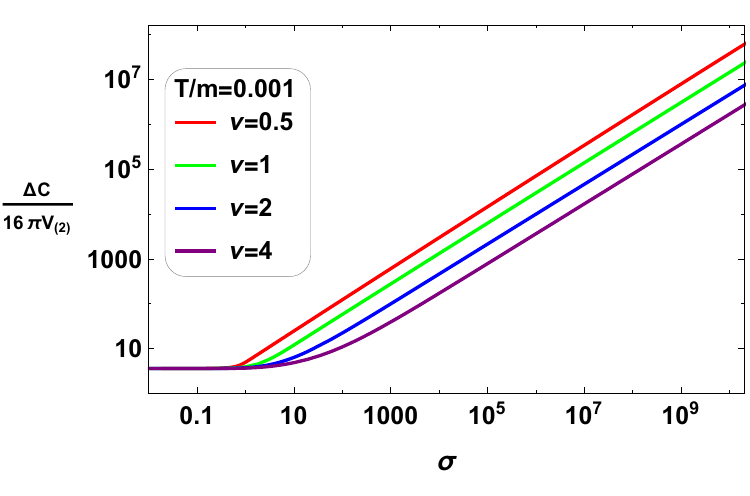}
    \caption{The complexity of formation as functions of the shear stress for different values of $\nu$. For large shear, the complexity follows the scaling behavior with a universal power $\zeta\simeq 2/3$. Here, we have fixed $T/m=0.001$.}
    \label{CoFwithstress}
\end{figure}

To avoid divergence caused by the infinite boundary, we need to regularize the complexity, i.e., introducing the complexity of formation \cite{Chapman:2016hwi}. From a physical perspective, what we are considering here is the additional complexity involved in forming the TFD state, compared to preparing each of the two individual CFTs in the vacuum, i.e.,
\begin{eqnarray}
    \Delta\mathcal{C}=16\pi\,\Delta \mathcal{V}=16\pi\, \left(\mathcal{V}_{\text{BH}}-\mathcal{V}_{\text{AdS}}\right),
\end{eqnarray}
where the volumes can be obtained from
\begin{eqnarray}
\mathcal{V}&=&2\int_{\mathcal{B}}\sqrt{|\hat{h}|}dudxdy=2V_{(2)}\int^{u_h}_{\varepsilon_0}\frac{du}{u^3\sqrt{f(u)}},
\end{eqnarray}
with $\hat{h}$ the determinant of the induced metric of the $t=0$ slice, $\varepsilon_0$ the UV cutoff and $V_{(2)}=\iint dxdy$ the area of the boundary system. Note that $V_{(2)}$ can be directly factorized out due to the fact that $\hat{h}$ is only $u-$dependent and $\text{det}(\gamma_{ij})$ is a constant.
Then, the volume difference can be expressed as
\begin{eqnarray}\label{cinte}
    \Delta \mathcal{V}
&=&2V_{(2)}\left[\int^{u_h}_{\varepsilon_0\rightarrow0}\frac{du}{u^3} \left(\frac{1}{\sqrt{f(u)}}-1\right)-\frac{1}{2u^2_h}\right],
\end{eqnarray}
where we have taken $f(u)=1$ for AdS vacuum and $\varepsilon_0\rightarrow0$ so that the integration covers the whole Cauchy surface. 
For different temperatures, one can plot $\Delta\mathcal{C}/(16\pi V_{(2)})$ as functions of the shear stress $\sigma$. The numeric results for a fixed temperature and different values of $\nu$ have been shown in Fig. \ref{CoFwithstress}. 

For all cases, we find that the complexity of formation grows as the increase of the shear stress and displays a scaling $
\Delta\mathcal{C}\sim \sigma^{\zeta}
$
for large shear stresses, in analogy with equation \eqref{enscaling} in the same regime. Nevertheless, one can check that the power  $\zeta$ takes a universal value, $2/3$, which is different from the power of the entropy that is dependent on  the details of the model ($\nu$, or the parameters $M$ and $N$).

In order to have a clearer picture about the numeric result, we shall look at the black hole geometry for large shear. When $\epsilon$ is large enough, a `near-horizon geometry' with a special scaling symmetry emerges  and its blackening factor is given by \cite{Baggioli:2020qdg}:
\begin{align}
     f(u)\simeq f_0\left[1-\left(\frac{u}{u_h}\right)^{3+\frac{9}{\nu^2}}\right],\    \ f_0=\frac{1}{(1+\frac{3}{\nu M})(1+\frac{3}{\nu^2})}. \label{scalinggeo}
\end{align} 
This geometry is correct from $u_h$ up to a UV scale $u_*$ that is controlled by the shear strain. Such an emergent scale $u_*$
provides the gravity dual of a Lifshitz-like fixed point. Interestingly, when $\epsilon \rightarrow \infty$, it is found that $u_*\sim \epsilon^{-\nu/3}\rightarrow 0$, and such an region expands towards the boundary. Eventually, the full bulk solution can be approximated by the this geometry. Note that the existence of such a solution
depends (but only slightly) on the choice of the function $V(X,Z)$.
A brief revisit of this emergent geometry can be found in the Appendix.

Then, the integral in equation \eqref{cinte} can be carried out analytically, yielding
\begin{eqnarray}\label{intgrationlarge}
\Delta\mathcal{C}&\simeq& 16\pi V_{(2)}\left(\frac{1}{\sqrt{f_0}}-1\right)u_*^{-2}+\dots, \label{DeltaCR}
\end{eqnarray}
where the term proportional to $u_h^{-2}$ has been neglected since $u_*\ll u_h$ when $\epsilon \gg 1$. The detailed derivation has been shown in the Appendix. On the other hand, from the UV expansion  of $\gamma_{xy}$, the shear stress scales like $\sigma\sim u_*^{-3}$ in the same regime.
Combining this and equation \eqref{intgrationlarge}, we obtain the universal scaling $\Delta\mathcal{C}\sim \sigma^{2/3}$ exactly. Therefore, from the holographic perspective, the emergence of the universal scaling in FIG. \ref{CoFwithstress2} is not a coincidence but a result from  the fact that shear stress and complexity are both controlled by the same scale $u_*$ under large shear. 

The situation for thermal entropy, however, is slightly different since it is determined by another scale, the size of the horizon $u_h$. 
Although $u_*$ and $u_h$ are related to each other, their relationship highly depends on the details of systems that are characterized by $M$ and $N$. This is why the value of the scaling power $\xi$ in \eqref{enscaling} is not universal. For the same reason, our findings also provides a counterexample to the previous understanding that the complexity of formation is proportional to the Bekenstein-Hawking entropy for large (or high temperature) static black holes \cite{Chapman:2016hwi,Chapman:2017rqy,Carmi:2017jqz}. Since $u_h\sim\epsilon^{-\nu/\left(3+\nu^2/3\right)}$, a large separation between $u_*$ and $u_h$ arises in the scaling regime, resulting in a big deviation of $\Delta \mathcal{C}/V_{(2)}$ from $s$. However, due to $u_*\ll u_h$, the complexity is still bounded from below by the thermal entropy of the black hole \cite{AlBalushi:2020rqe,AlBalushi:2020heq}. In FIG. \ref{ratiocds2} in the Appendix, we show how the shear strain affects the relationship between entropy density and complexity. Indeed, it is found that $\Delta \mathcal{C}/V_{(2)} $ deviates significantly from $s$ when the strain is sufficiently large.
\\

\section{Generalization in higher dimensions} 

In the previous two sections, the investigation is only restricted to two spatial dimensions. Now, let us generalize our conclusion to higher dimensions. In higher-dimensional materials, the mode of shear can be more diverse. The most straightforward extension of the shear in 2D systems is to consider the `simple shear deformations' for higher-dimensional systems, i.e., considering anisotropic effects only within a plane. As a result, the simple shear just reduces to a 2D case, which makes the analysis much easier.
Note that this has been verified in simulations of a specific 3D amorphous system \cite{Pan:2021cux}. 

When loading pure shear deformations in the $x-y$ plane of a higher dimensional system, we need to introduce more axions in the bulk and choose their profiles as follows:
\begin{eqnarray}\label{shearprohighD}
&M^{1}_{1}=M^{2}_{2}=\sqrt{1+\epsilon^2/4},\quad M^{1}_{2}=M^{2}_{1}=\epsilon,\\ \nonumber
&M^{I}_{i}=\delta^{I}_{i},\quad I,i=3,4,\dots, d.
\end{eqnarray}
Since the axions only break the rotations in the $x-y$ plane, one can expect that, for very large shear, the anisotropic geometry with Lifshitz like scaling appears again . Note that such kind of geometries are actually not uncommon and had already been discovered extensively in (4+1)-dimensional holographic systems \cite{Mateos:2011ix,Mateos:2011tv,Rebhan:2011vd,Landsteiner:2015pdh,Landsteiner:2016stv}. For $d\ge 2$, equation \eqref{cinte} should be replaced with: 
\begin{align}
    \Delta \mathcal{V}
&=&2\,V_{(d)}\left[\int^{u_h}_{\varepsilon_0}\frac{du}{u^{d+1}} \left(\frac{1}{\sqrt{f(u)}}-1\right)-\frac{1}{d\,u^d_h}\right], \label{volumed}
\end{align}
which now leads to $\Delta\mathcal{C}\sim u_*^{-d}$. Additionally, we know that $\gamma_{ij}$ behaves like \footnote{When $d$ is odd, a
logarithmic term $\sim u^{d+1}\text{Log}(u)$ appears in the UV expansion. This term contributes a non-trivial Weyl anomaly on boundary. It should be interesting to investigate the shear effects on the Weyl anomaly within our holographic setup which is, however, beyond the scope of this work.}:
\begin{eqnarray}\label{generaldUV}
\gamma_{ij}(u)\sim \gamma_{ij}^{(0)}+\dots+\gamma_{ij}^{(d+1)}u^{d+1}+\dots,
\end{eqnarray}
near the boundary, implying that $\sigma \sim u_*^{-(d+1)}$. Combining these, we then obtain the universal relation \eqref{scalingintro} for $(d+1)$-dimensional systems under simple shear.

\section{Discussion and outlook}
The holographic computations reveal a direct connection between nonlinear elasticity and quantum complexity in amorphous solids under shear. In the limit of large shear deformations, shear stress and complexity of formation obey a universal scaling relation with a power independent of the microscopic details. At first glance, this connection seems quite unexpected, since it links a local observable (one-point correlator) with a non-local one. However, from the bulk perspective, such a connection naturally arises from the dependence of the two observables on the same emergent scale $u_*$, which bounds the scaling geometry from the UV side. This is reminiscent of the well-known example in the holographic study of isotropic fluids, where the shear viscosity $\eta$ and the entropy density $s$ are both governed by the horizon scale $u_h$, leading to a universal ratio $\eta/s = 1/4\pi$ \cite{Kovtun:2004de}.

This finding also enhances our knowledge of black hole physics, indicating that complexity is generally not directly controlled by the thermal entropy. Despite the fact that such a statement was firstly made in the study of rotating black holes \cite{AlBalushi:2020rqe,AlBalushi:2020heq}, we have extended it to the case of static black holes in this work. Furthermore, since shear deformations do not bring any modifications to the definition of the thermodynamic volume, our results should not follow the ``complexity-thermodynamic volume connection" proposed in \cite{AlBalushi:2020rqe,AlBalushi:2020heq}.

Nevertheless, a question naturally arises: Is the connection between nonlinear elasticity and complexity in amorphous materials testable by experiments, simulations, or other theoretical approaches? Unfortunately, we do not yet have a complete understanding about the physical nature of the complexity within the CV conjecture. However, recent developments have revealed some similarities between the CV conjecture and specific state and operator complexities, particularly the Krylov complexity, making future verification possible \cite{Chapman:2017rqy,Belin:2021bga,Belin:2022xmt,Myers:2024vve,Adhikari:2022whf,Zhai:2024tkz,Sanchez-Garrido:2024pcy,Heller:2024ldz}. Finally, as a parallel study, it would also be intriguing to calculate the complexity within the framework of the complexity=action (CA) conjecture \cite{Brown:2015bva} and check whether the associated complexity is still governed by the emergent UV scale. \\
 
%######################################################################%
\section*{Acknowledgments}
We would like to thank Matteo Baggioli, Wen-Cong Gan, Xiao-Mei Kuang, Li Li and Run-Qiu Yang for fruitful discussions and helpful comments. This work is supported by the National Natural Science Foundation of China (NSFC) under Grants No.12275038 and No.12447178.

\bibliography{ref-lib}

\clearpage %
%\renewcommand{\thesubsection}{{a.\arabic{subsection}}}
%\renewcommand{\thesection}{\Alph{section}}
%\setcounter{section}{0}
%\titleformat*{\section}{\centering \bfseries}
%\onecolumngrid

\renewcommand{\thesubsection}{{\alph{subsection}}}
\setcounter{section}{0}
\titleformat*{\section}{\centering \Large \bfseries}

\onecolumngrid
%\appendix

\section*{Appendix}
\subsection{Numeric solutions under finite shear deformations}
The general asymptotically AdS background that is anisotropic in $x-y$ plane takes the following form:
\begin{eqnarray}
    ds^2=\frac{L^2}{u^2}\left[-f(u)e^{-\chi(u)}dt^2+\frac{1}{f(u)}du^2+\gamma_{ij}(u)dx^{i}dx^j\right], \quad i,j=x,y \label{metic}
\end{eqnarray}
where $\gamma_{ij}(u)=\delta_{ij}$ in the absence of shear deformation. To facilitate numerical calculations, we express $\gamma_{ij}(u)$ and $M^{I}_{i}$ in terms of the following hyperbolic functions in the general case:
\begin{eqnarray}
    \gamma_{ij}(u)=\begin{pmatrix}
        \cosh[h(u)]  &  \sinh[h(u)] \\ 
        \sinh[h(u)]  &  \cosh[h(u)]
    \end{pmatrix}.
\end{eqnarray}
Similarly, using the transformation relation for the background shear strain $\epsilon=2\sinh\left(\frac{\Omega}{2}\right)$, the matrix form 
\eqref{shearpro} of the axion field can be rewritten as: 
\begin{eqnarray}
     M^{I}_{i}=\alpha\begin{pmatrix}
    \cosh(\Omega/2)  &  \sinh(\Omega/2) \\
     \sinh(\Omega/2) &   \cosh(\Omega/2)
\end{pmatrix}.
\end{eqnarray}
This formulation ensures consistency with the conventions used in the context of anisotropic AdS backgrounds and is suitable for numerical analysis.
Then, the equations of motion can be written as 
\begin{eqnarray}
    0&=&2\chi'-uh'^2,    \label{EOM1}\\
    0&=&f(u\chi'+6)-2(uf'+3-m^2V(\bar{X},\bar{Z})),  \label{EOM2}\\
    0&=&h''+h'\left(\frac{f'}{f}-\frac{2}{u}\right)-\frac{1}{4}uh'^3-\frac{2m^2V_h(\bar{X},\bar{Z})}{u^2f(u)},\label{EOM3}
\end{eqnarray}
where $\bar{X}=\alpha^2u^2\cosh(\Omega-h(u))$ and $\bar{Z}=\alpha^4u^4$ represent the background values of $X$ and $Z$, respectively.
The Hawking temperature and the entropy density are given by:
\begin{eqnarray}
    T=-\frac{f'(u_h)e^{-\chi(u_h)/2}}{4\pi}=\frac{3-m^2V(\bar{X}_h,\bar{Z}_h)}{4\pi u_h}e^{-\chi(u_h)/2},\quad  s=\frac{4\pi}{u^2_h},\label{therm}
\end{eqnarray}
with $\bar{X}_h$ and $\bar{Z}_h$ denoting the background values at the horizon $u_h$, respectively.
To solve the equations \eqref{EOM1}$-$\eqref{EOM3}, it is necessary to impose boundary conditions at the horizon and near the UV boundary. At the black hole horizon, we have $f(u_h)=0$ and $h(u_h)=h_h$, where $h_h$ is a constant. At the UV boundary, we require that $f(0)=1$ and $\chi(0)=0$. The boundary condition $\gamma_{xy}^{(0)}=0$ means $h(0)=0$. Furthermore, by fixing $\alpha=1$, the numerical solutions to the equations corresponding to different values of $h_{h}$ can be obtained using the shooting method.

The asymptotic  behavior of $h$ near the boundary is given by:
\begin{eqnarray}
h(u)=\mathcal{H}_0+\mathcal{H}_3u^3+\dots.
\end{eqnarray}
Then, the boundary condition $h(0)=0$ can be realized by fixing the leading term $\mathcal{H}_0=0$. According to the holographic dictionary, 
the shear stress can be extracted as \footnote{Note that some previous papers adopted a different convention $8\pi G_N\equiv1$. That why there is an extra $1/2$ in those papers.}:
\begin{eqnarray}
    \sigma\equiv \left\langle T_{xy} \right\rangle=3\gamma_{xy}^{(3)}=3\mathcal{H}_3.
\end{eqnarray}
It is worth noting that when $\mathcal{H}_0\neq 0$, the gravity sector also provides an additional source for the boundary stress tensor. In this case, the relationship between $\sigma$ and $\mathcal{H}_{0,3}$ becomes more complex, and the parameter $\epsilon$ (or $\Omega$) can no longer be interpreted solely as the shear strain.

\subsection{Anisotropic Lifshitz like geometry for large shear strain}
\begin{figure}
    \centering
    \includegraphics[width=0.45
    \linewidth]{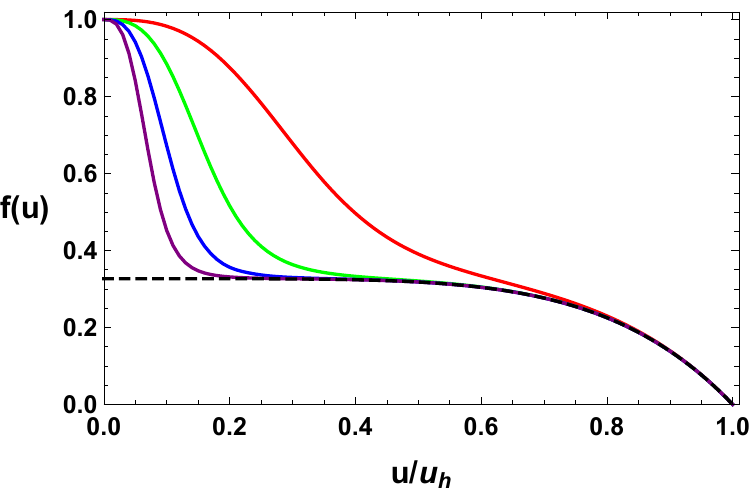}
\includegraphics[width=0.45
    \linewidth]{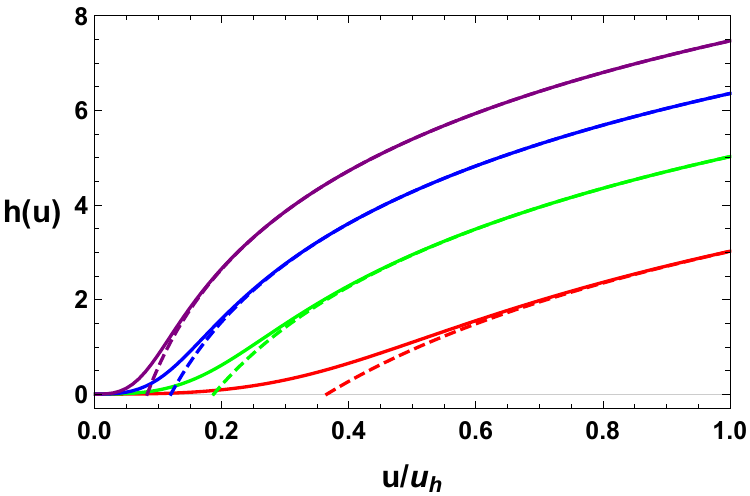}
\includegraphics[width=0.45
    \linewidth]{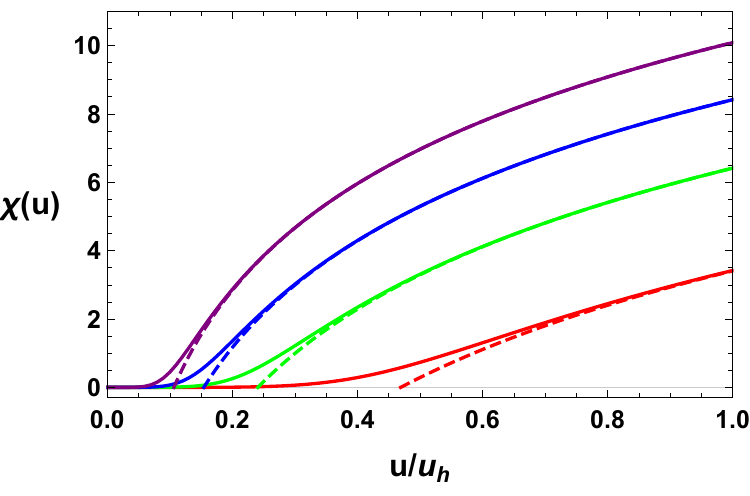}
    \caption{The functions $f(u)$, $h(u)$, and $\chi(u)$  for the model parameters $M=2, N=1/2$ at a fixed temperature $T/m=0.1$ for the different shear strains (i.e., $\epsilon\approx5.5\times10$(red), $\epsilon\approx1.4\times10^3$(green), $\epsilon\approx1.2\times10^4$(blue),$\epsilon\approx7.6\times10^4$(purple) ). The dashed line represents the scaling geometry \eqref{analyticgeo}. From the figure, it is evident that as the shear stress increases, the geometry gradually approaches the analytical scaling geometry. Here, $c_0\approx-1.11922$. }
    \label{blackfactor}
\end{figure}

This section provides a brief revisit of the Lifshitz like ``near horizon geometry" discovered in previous works \cite{Mateos:2011ix,Mateos:2011tv,Rebhan:2011vd,Baggioli:2020qdg}. With an appropriate choice of the function $V(X,Z)$, a special geometry appears when the shear strain becomes large. From the horizon $u_h$ up to an emergent scale $u_*$, which is controlled by $\epsilon$, the background solution can be expressed analytically as follows:
\begin{align}
   & f(u)\simeq f_0\left[1-\left(\frac{u}{u_h}\right)^{3+\frac{9}{\nu^2}}\right],\quad f_0=\frac{1}{(1+\frac{3}{\nu M})(1+\frac{3}{\nu^2})}<1,\nonumber\\
 & h(u)\simeq\frac{6}{\nu}\log\left(\frac{u}{u_*}\right),\quad
    \chi(u)\simeq\frac{18}{\nu^2}\log\left(\frac{u}{u_*}\right)+c_0, 
\label{analyticgeo}
\end{align}
where the constant $c_0$ can be determined by the first relation in \eqref{therm}.
%nserting the results above, we obtain that
% \begin{eqnarray}
%   ds^2&\simeq& \frac{1}{u^2}\left\{\frac{du^2}{f_0}-f_0\left(\frac{u}{u_*}\right)^{-\frac{18}{\nu^2}}dt^2+\frac{1}{2} 
%  \left[\left(\frac{u}{u_*}\right)^{-\frac{6}{\nu}}+\left(\frac{u}{u_*}\right)^{\frac{6}{\nu}}  \right]dx^2+\frac{1}{2} \left[\left(\frac{u}{u_*}\right)^{-\frac{6}{\nu}}+\left(\frac{u}{u_*}\right)^{\frac{6}{\nu}}  \right]dy^2 \right \nonumber\\
%    &&\left-\left[\left(\frac{u}{u_*}\right)^{-\frac{6}{\nu}}-\left(\frac{u}{u_*}\right)^{\frac{6}{\nu}}  \right] dxdy  \right\}
% \end{eqnarray}
The full numerical solution has been shown in Fig. \ref{blackfactor}. When $\epsilon \gg1$, we find that $u_*$ is significantly pushed towards the boundary. Consequently, the ``near horizon geometry" dominates the bulk solution. 

Finally, by rotating to the diagonal basis and absorbing the scale $u_*$, we find a Lifshitz like fixed point, which is dual to the following scaling solution:
\begin{equation}
    ds^2\simeq \frac{1}{u^2}\left[\frac{du^2}{f_0}-f_0u^{-\frac{18}{\nu^2 }}dt^2+ u^{-\frac{6}{\nu}}d\tilde{x}^2 +u^{\frac{6}{\nu}}d\tilde{y}^2  \right].
\end{equation}
This solution holds at  $u\simeq u_*$. It is important to note that this should be understood as an UV fixed point, as it acts as an attractor solution towards the UV and is significantly separated from the horizon.

\subsection{Computation of holographic complexity}

The complexity=volume (CV) conjecture states that the complexity of boundary states corresponds to the volume of a maximal constant-time slice in the bulk \cite{Stanford:2014jda}: 
\begin{eqnarray}   \mathcal{C}_{V}=\frac{\text{max}[\mathcal{V}(\mathcal{B})]}{G_NL}.
\end{eqnarray}
Here, we consider thermal field double states, which can be described by two-sided eternal black holes. With the convention $16\pi G_N\equiv 1$ and $L\equiv1$,
the complexity of formation is calculated using the following subtraction scheme:
\begin{eqnarray}
\Delta\mathcal{C}=16\pi\,\Delta \mathcal{V}=16\pi\, \left(\mathcal{V}_{\text{BH}}-\mathcal{V}_{\text{AdS}}\right). \label{DeltaC}
\end{eqnarray}
For $t_L=t_R=0$, the time slice of maximal volume in the bulk corresponds to a straight line passing through the bifurcation surface in Fig. \ref{ERB}.
For the black hole and pure AdS vacuum, the volumes of the associated time slices can be calculated as:
\begin{eqnarray} \mathcal{V}_{\text{BH}}&=&2\int \sqrt{|\hat{h}_\text{BH}|}\,dudx^1\dots dx^d=2V_{(d)}\int^{u_h}_{\varepsilon_0}\frac{du}{u^{d+1}\sqrt{f(u)}},\\
\mathcal{V}_{\text{AdS}}&=&2\int \sqrt{|\hat{h}_{\text{AdS}}|}\,dudx^1\dots dx^d=2V_{(d)}\int^{\infty}_{\varepsilon_0}\frac{du}{u^{d+1}},
\end{eqnarray}
where $\varepsilon_0$ is the UV cutoff, $V_{(d)}=\int dx^1\dots dx^d$ and we have used the condition $\text{det}(\gamma_{ij})=1$ for the pure shear case. From these, we derive:
\begin{eqnarray}
    \Delta \mathcal{V}=\mathcal{V}_{\text{BH}}-\mathcal{V}_{\text{AdS}}
&=&2V_{(d)}\left[\int^{u_h}_{\varepsilon_0}\frac{du}{u^{d+1} \sqrt{f(u)}}-\int^{\infty}_{\varepsilon_0}\frac{du}{u^{d+1}}\right] \nonumber\\
&=&2V_{(d)}\left[\int^{u_h}_{\varepsilon_0}\frac{du}{u^{d+1}} \left(\frac{1}{\sqrt{f(u)}}-1\right)-\frac{1}{d\,u^{d}_h}\right],
\end{eqnarray}
which corresponds to equation \eqref{volumed} in the main text.

\subsection{Stress, entropy and complexity under finite shear deformations}

\begin{figure}
    \centering
\includegraphics[width=0.45
    \linewidth]{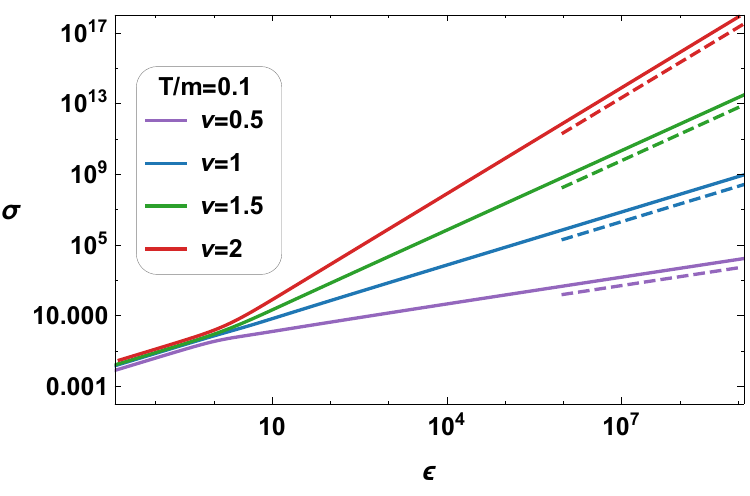}
\includegraphics[width=0.44
    \linewidth]{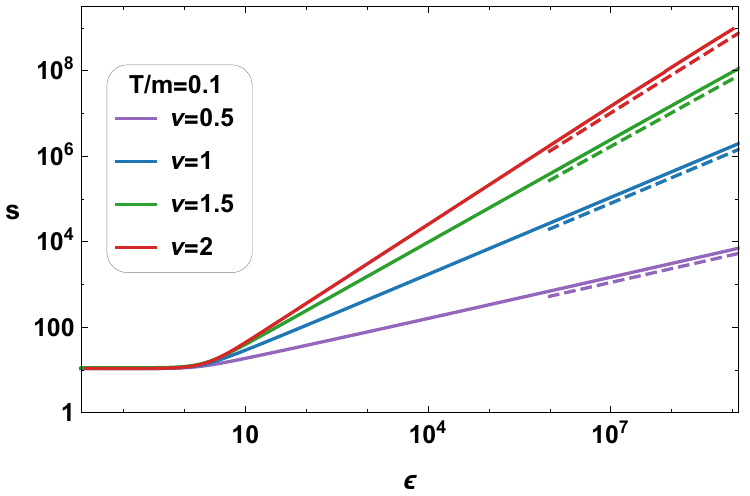}
\includegraphics[width=0.46
    \linewidth]{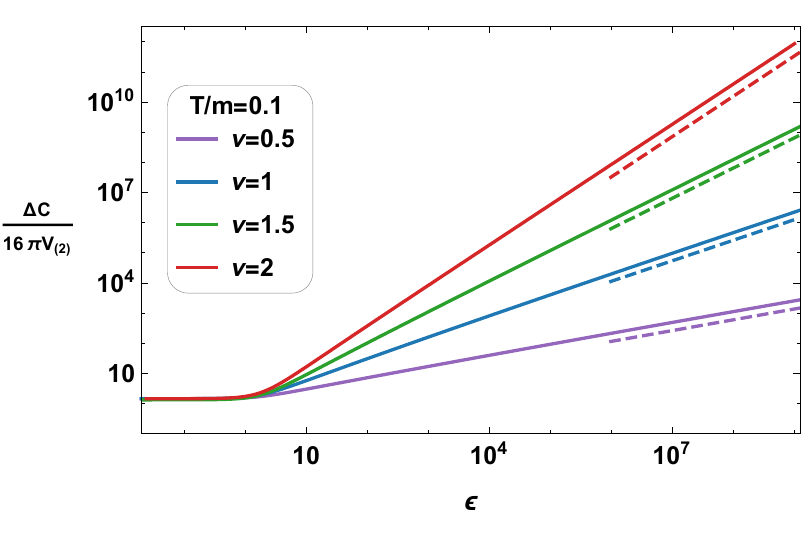}
    \caption{Shear stress, entropy density, and complexity of formation as functions of the shear strain for different values of $\nu$.}
    \label{scalingvsepsilon}
\end{figure}

\begin{figure}
    \centering
\includegraphics[width=0.55
    \linewidth]{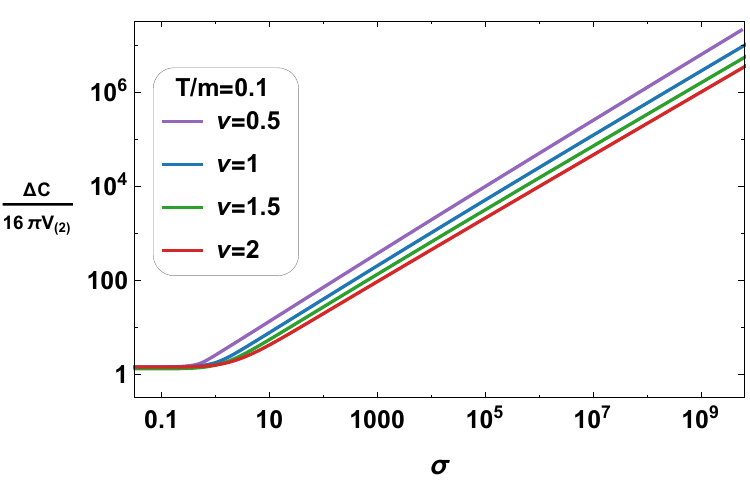}
    \caption{Complexity of formation as a function of shear stress for different values of $\nu$. For large shear, the results exhibit the same scaling behavior as shown in Fig. \ref{CoFwithstress} for $T/m=0.001$.}
    \label{CoFwithstress2}
\end{figure}

For $d=2$, the evolutions of the shear stress, the thermal entropy, and the complexity under finite shear have been displayed in FIG. \ref{scalingvsepsilon}. Additionally, it is found that $\Delta \mathcal{C}\sim \sigma ^{2/3}$ which has been shown in Fig. \ref{CoFwithstress} and Fig. \ref{CoFwithstress2} for $T/m=0.001$ and $T/m=0.1$. 

In the very large shear regime, the ``near horizon" geometry \eqref{analyticgeo} dominates in the bulk. Consequently, the volume of the time slice can be calculated analytically as follows:
\begin{eqnarray}
    \Delta \mathcal{V}
&\approx&V_{(2)}\left\{ \left(\frac{1}{u^2_h}-\frac{1}{u^2_*}\right)-\frac{1}{u^2_h}
+\frac{1}{\sqrt{f_0}}\left[\frac{1}{u^2_*}-\frac{\sqrt{\pi}\Gamma\left(\frac{1}{3}+\frac{2}{3+\nu^2}\right)}{\Gamma\left(-\frac{1}{6}+\frac{2}{3+\nu^2} \right)  }\frac{1}{u^2_h}\right]    \right\}  \nonumber\\
&=&V_{(2)}\left[\left(\frac{1}{\sqrt{f_0}}-1\right)\frac{1}{u^2_*}-\frac{\sqrt{\pi}\Gamma\left(
\frac{1}{3}+\frac{2}{3+\nu^2}\right)}{\sqrt{f_0}\Gamma\left(-\frac{1}{6}+\frac{2}{3+\nu^2} \right)  }\frac{1}{u^2_h}  \right]\nonumber\\
&\approx&V_{(2)}\left(\frac{1}{\sqrt{f_0}}-1\right)u^{-2}_*+\dots.\label{volumesim}
\end{eqnarray}
In the last step above, we have neglected the other term since $u_*\ll u_h$. 

\begin{figure}
    \centering
\includegraphics[width=0.55
    \linewidth]{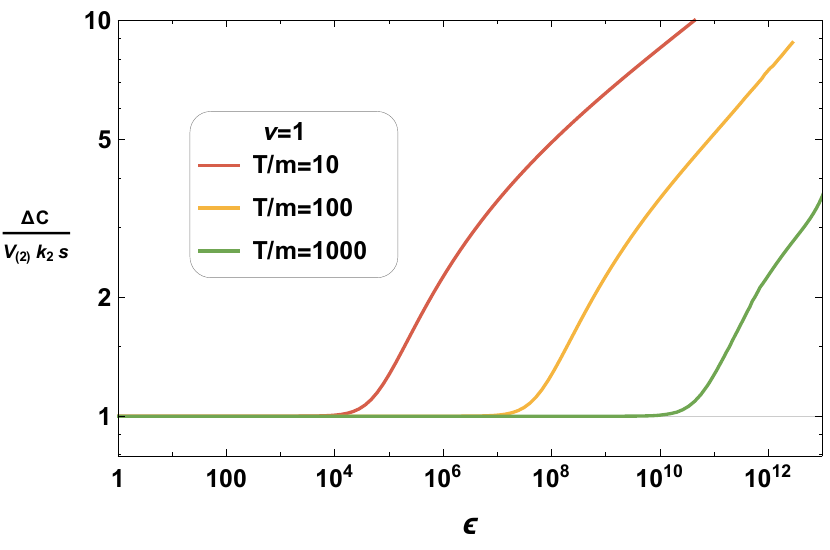}
    \caption{The ratio $\Delta \mathcal{C}/(V_{(2)} k_{2} s)$ as a function of $\epsilon$ for different temperatures $T/m$. Here, we have fixed $\nu=1$. The horizontal gray line denotes the value from the ``complexity-entropy connection". It is found that, regardless of the size or temperature of the black hole, the proposed connection always breaks down when the strain is sufficiently large.}
    \label{ratiocds2}
\end{figure}

On the other hand, for $d+2$-dimensional black holes, it was proposed that the complexity scales linearly with the entropy \cite{Chapman:2016hwi}: 
\begin{eqnarray}
\frac{\Delta \mathcal{C}}{ V_{(d)}}=4\sqrt{\pi}\frac{(d-1)\,\Gamma\left(1+\frac{1}{d+1}\right)}{d\,\Gamma\left(\frac{1}{2}+\frac{1}{d+1}\right)}s\equiv k_d s. \label{proposal}
\end{eqnarray}
However, comparing \eqref{therm} and our result \eqref{volumesim}, this proposal is no longer valid for sheared static black holes. In Fig. \ref{ratiocds2}, we explicitly demonstrate how the proposal \eqref{proposal} breaks down as the shear strain is enhanced. In the very large shear regime, one can expect that $\Delta \mathcal{C}/(V_{(2)} k_{2} s)\gg1$. Furthermore, for all cases, the complexity is bounded from below by the entropy. 

%\begin{figure}
    %\centering
%\includegraphics[width=0.55
  %  \linewidth]{cdsvstm1.pdf}
 %   \caption{The ratio $\Delta \mathcal{C}/(  V_{(2)} k_{2} s  )$ as a function of $T/m$ for different strains. Here, we have fixed $\nu=2$.}
   % \label{ratiocds1}
%\end{figure}

\newpage

 \end{document}